\documentclass[%
 reprint,
superscriptaddress,
 amsmath,amssymb,
 aps,
prl,
]{revtex4-2}
\usepackage{mathrsfs}
\usepackage{orcidlink}
\hypersetup{
     colorlinks=true,  
     linkcolor=blue,   
     citecolor=blue,
     urlcolor=blue
}
\usepackage{graphicx}
\usepackage{dcolumn}
\usepackage{bm}
\usepackage{empheq}

\renewcommand{\d}{\mathrm{d}}
\begin{document}

\preprint{APS/123-QED}

\title{Non-Maxwellianity in Temperature Relaxation between Plasma Species}

\author{Yao Guo\orcidlink{0009-0009-7523-5887}}
\affiliation{%
State Key Laboratory of Dark Matter Physics, Key Laboratory for Laser Plasmas, Department of Physics and Astronomy, Shanghai Jiao Tong University, Shanghai 200240, People's Republic of China
}%

\author{Dong Wu\orcidlink{0000-0001-5738-5739}}%
 \email{dwu.phys@sjtu.edu.cn}

\affiliation{%
State Key Laboratory of Dark Matter Physics, Key Laboratory for Laser Plasmas, Department of Physics and Astronomy, Shanghai Jiao Tong University, Shanghai 200240, People's Republic of China
}%
\affiliation{Collaborative Innovation Center of IFSA (CICIFSA), Shanghai Jiao Tong University, Shanghai 200240, People's Republic of China}

 \author{Jie Zhang\orcidlink{0000-0001-7821-4808}}
 \email{jzhang1@sjtu.edu.cn}
 
\affiliation{%
State Key Laboratory of Dark Matter Physics, Key Laboratory for Laser Plasmas, Department of Physics and Astronomy, Shanghai Jiao Tong University, Shanghai 200240, People's Republic of China
}%

\affiliation{Collaborative Innovation Center of IFSA (CICIFSA), Shanghai Jiao Tong University, Shanghai 200240, People's Republic of China}

\affiliation{Institute of Physics, Chinese Academy of Sciences, Beijing 100190, People's Republic of China}

\date{\today}

\begin{abstract}
Temperature relaxation between plasma species is usually described by Landau--Spitzer (LS) theory, which assumes Maxwellian distributions throughout. We show analytically that relaxation between two initially Maxwellian species in homogeneous isotropic plasmas generates a universal three-lobe distortion that reduces energy exchange over a broad parameter range. Full Fokker--Planck simulations confirm these predictions and reveal substantial relaxation delays, most pronounced at intermediate mass ratios. For a post-shock DT fusion plasma with $T_{\mathrm{T}}/T_{\mathrm{D}}=1.5$, relaxation takes 47\% longer than the LS prediction, while the distortions lower the apparent DD/DT temperature ratio and produce positive DD and negative DT Gamow-shift residuals. These results also have implications for $\alpha$-driven fuel heating and equilibration in astrophysical plasmas.
\end{abstract}

\maketitle


\textit{Introduction}---Temperature relaxation between plasma species is a fundamental problem in plasma physics~\cite{landau1936kinetische,spitzer1956physics,Sprenkle2022}. Its standard weak-coupling description is the Landau--Spitzer (LS) equation,
\begin{align}
    \frac{\mathrm{d}T_a}{\mathrm{d}t} &= \nu^E_{ab}(T_b - T_a),  \nonumber \\ 
    \nu^E_{ab} &= \frac{8\sqrt{2\pi} n_b Z_a^2 Z_b^2 e^4 \ln\Lambda_{ab}}{3m_am_b} \left( \frac{T_a}{m_a} + \frac{T_b}{m_b} \right)^{-3/2}, \label{eq:spitzer}
\end{align}
which is derived by taking the energy moment of the Landau operator while constraining each species to remain Maxwellian. This constraint is well justified for Landau's electron--ion setting \cite{landau1936kinetische}, where self-thermalization is much faster than interspecies energy exchange. However, it is also widely used for relaxation between arbitrary species \cite{Braginskii1958,Braginskii1965,TAKIZUKA1977205,Nanbu1997,Dimonte2008,TAITANO2015357,Faussurier2016}.

Non-Maxwellian corrections to Eq.~(\ref{eq:spitzer}) have been derived for electron--ion relaxation near thermal equilibrium \cite{Gorev2015} and for electrons interacting with much hotter ions, where the low-energy part of the electron distribution is distorted~\cite{rosenbluth1976energy,Todd1995,Bobylev1997}. For comparable ion species, interspecies exchange should compete with self-thermalization over a broader velocity range. Indeed, clear deviations from the LS equation have been observed for ions in kinetic and molecular dynamics simulations \cite{Rambo1995,Nanbu1997,Benedict2012,MEISL2013S342,Bellei2014}, but a systematic understanding of how interspecies collisions deform the distributions and thereby modify temperature relaxation is still lacking.

In this Letter, we identify how collisional relaxation generates distribution distortions that feed back on interspecies energy exchange.
Starting from two Maxwellian species at unequal temperatures, we derive the universal three-lobe structure of the initial deformation and establish a general criterion for its initial suppression of the LS energy-exchange rate.
Full Fokker--Planck calculations reveal substantial relaxation delays, most pronounced at intermediate mass ratios.
For representative DT plasmas, the same distortions both delay equilibration and modify fusion neutron spectra. These findings also suggest that collisional shape feedback may affect fuel heating by fusion \(\alpha\) particles.

\begin{figure*}
\includegraphics[width=1.98 \columnwidth]{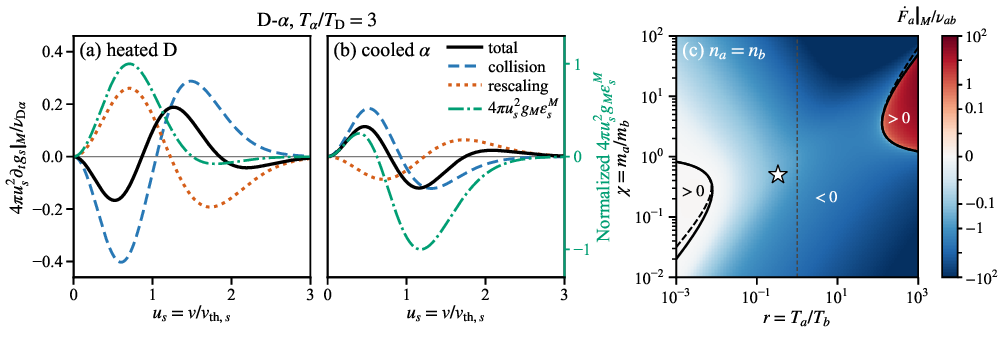}
\caption{Initial distribution deformation and feedback on energy exchange.
(a) and (b) Deformation sources
$4\pi u_s^2\partial_t g_s|_M/\nu_{\mathrm{D}\alpha}$
for D and $\alpha$, respectively, at $T_\alpha/T_{\mathrm D}=3$.
Contributions from interspecies collisions and thermal rescaling
are shown separately, together with the energy transfer rate per unit $u_s$, $4\pi u_s^2g_M\epsilon_s^M$, normalized by its maximum magnitude.
Here $\epsilon_s^M(u)\equiv\epsilon_s(u,g_M)$ denotes the normalized
single-particle energy transfer rate in a Maxwellian background.
(c) Map of $\dot F_a|_M/\nu_{ab}$ for $n_a=n_b$, with $r\equiv T_a/T_b$.
The solid curve marks $\dot F_a|_M=0$; the dashed
curve shows the corresponding boundary for $n_a/n_b=0.1$.
The star denotes the D--$\alpha$ case in panel (a). Red and blue indicate $\dot F_a|_M>0$ and
$\dot F_a|_M<0$, respectively.}
\label{fig:Fdot}
\end{figure*}

\textit{Initial departure from Maxwellians}---We consider two charged species \(a\) and \(b\) with spatially homogeneous and isotropic distributions; when both are ions, their coupling to electrons is neglected unless stated otherwise. For species $s$, we define $v_{\mathrm{th},s}=\sqrt{2T_s/m_s}$, $u_s=v/v_{\mathrm{th},s}$, and a distribution shape function $g_s(u_s)= f_s(v) v_{\mathrm{th},s}^3/n_s$. A Maxwellian is $g_M(u)=\pi^{-3/2}\exp(-u^2)$.  $g_s(u)$ is constrained by $\int_0^\infty 4\pi u^2 g_s(u)\,\d u=1$ and $\langle u^2\rangle\equiv\int_0^\infty 4\pi u^4 g_s(u)\,\d u=3/2$. In terms of $g(u)$, the kinetic equation for species $a$ is
\begin{equation}
 \frac{\partial g_a}{\partial t}-\frac{\dot T_a}{2T_a}\hat Dg_a=\hat C_{ab}[g_a,g_b]+\hat C_{aa}[g_a,g_a], \label{eq:FP2}
\end{equation}
where $\hat Dg\equiv3g+u\partial_ug$ accounts for thermal rescaling of the velocity and distribution amplitude as $v_{\mathrm{th}}$ evolves. The isotropic Rosenbluth--Fokker--Planck collision operator expressed in $u$ is \cite{Rosenbluth1957}
\begin{equation}
\hat C_{ab}[g_a,g_b]\!\!=\!\!\frac{4\pi\nu_{ab}}{u^2}\frac{\partial}{\partial u}\bigg[\chi\hat A_b(\alpha u)g_a
\!+\!\alpha u^2\hat B_b(\alpha u)\frac{\partial g_a}{\partial u}\bigg], \label{eq:RMJ}
\end{equation}
where $\hat A_b(x)=\int_0^{x}\zeta^2 g_b\,\d \zeta$ and $\hat B_b(x)=\int_0^{x}\zeta^4 g_b\,\d\zeta/(3x^3)+\int_{x}^\infty \zeta g_b\,\d\zeta/3$. Here we define $\alpha\equiv v_{\mathrm{th},a}/v_{\mathrm{th},b}$, $\chi\equiv m_a/m_b$, and $\nu_{ab}=4\pi n_bZ_a^2Z_b^2e^4 \ln \Lambda_{ab}/m_a^2v_{\mathrm{th},a}^3$. For simplicity, Coulomb logarithms $\ln\Lambda_{ab}$ are held fixed in this Letter unless stated otherwise.

The energy moment of Eq.~(\ref{eq:FP2}) is 
\begin{align}
\frac{3\dot{T}_a}{2T_a}&= \int_0^\infty4\pi u^4 \hat{C}_{ab}[g_a,g_b]\,\d u  \equiv Q_a[g_a,g_b] \nonumber\\
&=\int_0^\infty
4\pi u^2g_a(u)\,
\epsilon_a(u,g_b)\,\d u, \label{eq:Srate}
\end{align} 
where $\epsilon_a(u,g_b)=8\pi\nu_{ab}\left[
\alpha\int_{\alpha u}^{\infty}\zeta g_b(\zeta)\,\d\zeta
-\chi\hat A_b(\alpha u)/u\right]$ is the normalized single-particle energy transfer rate for $a$ at $u$. If both $g_a$ and $g_b$ are Maxwellian, Eq.~(\ref{eq:Srate}) exactly reduces to Eq.~(\ref{eq:spitzer}). However, $g_M$ itself is not an invariant solution of Eq.~(\ref{eq:FP2}). Let both species be Maxwellian at one instant. The self-collision term $\hat C_{aa}$ then vanishes, and Eq.~(\ref{eq:FP2}) gives
\begin{align}
    \frac{\partial g_a}{\partial t}\bigg|_M
&=g_M(u)\bigg\{(3-2u^2)\frac{\dot T_a}{2T_a}\bigg|_M+2\nu_{ab}(\chi-\alpha^{-2}) \nonumber \\
&\quad\times\left[\alpha(1+\alpha^2)\operatorname{erf}'(\alpha u)
-\frac{\operatorname{erf}(\alpha u)}{u}\right]\bigg\},
\label{eq:MaxwellianDeparture}
\end{align}
where the two terms arise from thermal rescaling and interspecies collisions, respectively. The erf terms are not proportional to $3-2u^2$, hence the right-hand side can vanish for all $u$ only when $\chi=\alpha^{-2}$, equivalently $T_a=T_b$. Therefore, for \(T_a\ne T_b\), interspecies relaxation immediately drives both species away from Maxwellians.

Figures~\ref{fig:Fdot}(a) and \ref{fig:Fdot}(b) show the deformation sources and their components for D--$\alpha$ with $T_\alpha/T_{\mathrm D}=3$. Both species exhibit a three-lobe distortion, but with opposite signs. In the End Matter, we prove that this three-lobe structure holds true for arbitrary pairs. It arises from the velocity dependence of Coulomb collisions: slow particles respond strongly to collisions, which control the low-$u$ deformation, whereas fast particles respond weakly, allowing thermal rescaling to dominate the high-$u$ tail. For heated D, collisions transfer low-velocity particles toward higher velocities, while the increasing thermal speed shifts the tail toward smaller $u_{\mathrm D}$. The resulting depletion at both ends produces a ``$-+-$'' source. For cooled $\alpha$, collisions redistribute particles toward lower velocities, while the decreasing thermal speed shifts the tail toward larger $u_\alpha$, producing the opposite ``$+-+$'' structure.

\textit{Feedback on relaxation rate}---The departure from Maxwellians immediately feeds back on
temperature relaxation. We define the relative correction to the relaxation rate as
\begin{equation}
F_a\equiv \frac{\dot T_a}{\dot T_a|_M}-1=
\frac{Q_a[g_a,g_b]}{Q_a^M}-1,
\label{eq:exactF}
\end{equation}
where $Q_a^M\equiv Q_a[g_M,g_M]$ is the LS energy moment evaluated at the same temperatures. In our two-species context, energy conservation gives $F_a=F_b\equiv F$. Starting from two Maxwellians, $F|_M=0$, while
\begin{align}
\dot F|_M&=
\frac{4\pi}{Q_a^M}\bigg[
\int_0^\infty u^2
\left.\partial_tg_a\right|_M
\epsilon_{a}(u,g_M)\, \d u
\nonumber  \\
&\quad-\alpha\chi
\frac{\nu_{ab}}{\nu_{ba}}
\int_0^\infty u^2
\left.\partial_tg_b\right|_M
\epsilon_{b}(u,g_M)\, \d u \bigg] \nonumber \\
&=\nu_{ab}{\cal P}(\alpha,\chi)
+\nu_{ba}{\cal P}(\alpha^{-1},\chi^{-1}).
\label{eq:exactFVariationEnd}
\end{align}
The explicit expression of ${\cal P}(\alpha,\chi)$ is given in the End Matter. Since $\nu_{ba}/\nu_{ab}
=(n_a/n_b)\chi^2\alpha^3$, the sign of $\dot F|_M$ depends only on $n_a/n_b$, $\chi$, and $\alpha$.  Figure~\ref{fig:Fdot}(c) maps this feedback for $n_a=n_b$. Except in the regions where the heavier species is much hotter than the lighter species, $\dot F|_M<0$. As shown by the analysis in the End Matter, we find that for arbitrary density and mass ratios with $T_a\ne T_b$,
\begin{equation}
\alpha_{hl}
<\alpha_c\simeq3.699
\quad\Longrightarrow\quad
\dot F|_M<0,
\label{eq:universalFdot}
\end{equation}
where $\alpha_{hl}$ is the thermal velocity ratio between the heavy and light species. Since $\alpha_{hl}\lesssim1$ in many relevant regimes, this initial feedback favors slower relaxation than predicted by LS theory.

This slowdown is a direct result of the distribution distortions shown in Figs.~\ref{fig:Fdot}(a) and \ref{fig:Fdot}(b). Since the part contributing most to the energy transfer rate $4\pi u_s^2g_M\epsilon_s^M$ is depleted for each species, $|Q_s|$ is decreased correspondingly. The exceptional region where $\dot F|_M>0$ is explained in the End Matter.

\textit{The relaxation process}---While Eq.~(\ref{eq:exactFVariationEnd}) determines the initial
trend, the subsequent relaxation depends on the accumulated feedback of the evolving distortions. We therefore investigate the relaxation using full FP simulations (see End Matter for numerical methods).

\begin{figure}[h]
\includegraphics[width=0.98\columnwidth]
{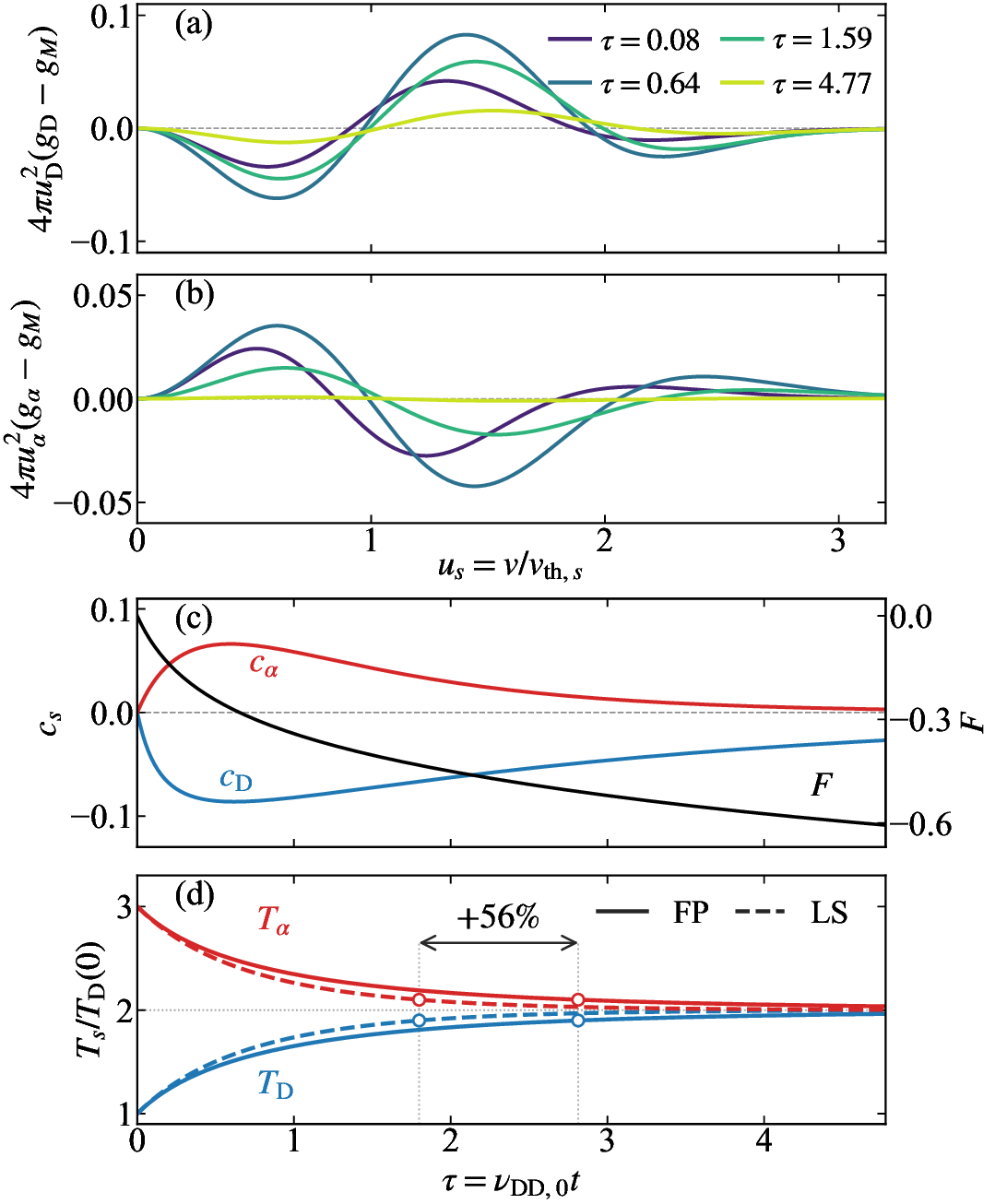}
\caption{Distribution deformation and delayed temperature relaxation.
(a) and (b) Deviations from $g_M$ for D and $\alpha$,
respectively, at $\tau=0.08,\,0.64,\,1.59,\,4.77$.
(c) Normalized fourth-moment departures $c_s$ and
rate correction $F$.
(d) D (blue) and $\alpha$ (red) temperatures from FP
(solid) and LS theory (dashed), normalized by $T_{\mathrm D}(0)$.
Circles mark a temperature difference of 10\% of its initial value.
The plasma is equimolar with $T_\alpha(0)/T_{\mathrm D}(0)=3$;
time is normalized as $\tau=\nu_{\mathrm{DD},0}t$.}
\label{fig:num}
\end{figure}

Figures~\ref{fig:num}(a) and \ref{fig:num}(b) show the
evolution of the distortions in an equimolar D--$\alpha$
plasma initialized with $T_\alpha/T_{\mathrm D}=3$.
Both species develop the predicted three-lobe structure,
which remains evident in the subsequent evolution.
We characterize the distortions by the lowest-order nontrivial moment, $\langle u_s^4\rangle$, with
\begin{equation}
c_s\equiv\frac{4}{15}\langle u_s^4\rangle-1,
\label{eq:shapeAmplitude}
\end{equation}
which is also the coefficient of $L_2^{(1/2)}(u^2)$
in the Sonine expansion of $g_s$.
Figure~\ref{fig:num}(c) shows $c_{\mathrm D}<0$ and
$c_\alpha>0$, consistent with their opposite distortions.
Their magnitudes grow, peak near $\tau\simeq0.6$,
and subsequently decay, while $|F|$ continues to increase,
reaching about $0.6$ at $\tau=4.77$.
Because $Q_a^M$ vanishes as $T_a\to T_b$, even weak distortions at small temperature differences can produce a substantial relative correction to the LS energy-exchange rate.

The accumulated feedback produces a substantial relaxation
delay [Fig.~\ref{fig:num}(d)].
Defining $t_{10}$ as the time at which the temperature
difference falls to 10\% of its initial value and
$\tau_{10}\equiv\nu_{\mathrm{DD},0}t_{10}$, we obtain
$\tau_{10}^{\mathrm{FP}}=2.813$ and
$\tau_{10}^{\mathrm{LS}}=1.798$, which means the kinetic relaxation takes approximately 56\% longer.
We note that for this two-species system,
$\tau_{10}^{\mathrm{FP}}-\tau_{10}^{\mathrm{LS}}
=-\int_0^{\tau_{10}^{\mathrm{FP}}}F\,\mathrm{d}\tau$.
For this case, retaining only the lowest nontrivial Sonine order,
$L_2^{(1/2)}(u^2)$, also reproduces the FP temperature history
(see End Matter).

\textit{Species parameter dependence}---The departure from
Maxwellian relaxation depends on how effectively interspecies
collisions deform each distribution relative to self-thermalization.
For a Maxwellian pair, we compare the fourth-moment source
$s_s\equiv\dot c_s|_M$ with the projected self-collision
damping rate of the $L_2^{(1/2)}$ perturbation,
$\gamma_{s2}=(8/15)\sqrt{2/\pi}\,\nu_{ss}$.
With $r= T_a/T_b$ and equal Coulomb logarithms,
these ratios are (see End Matter)
\begin{equation}
\begin{aligned}
{\cal W}_a\equiv\frac{|s_a|}{\gamma_{a2}}
&=\frac{3\sqrt2}{\eta}\,
\frac{\chi r^{3/2}|r-1|}{(r+\chi)^{5/2}},\\
{\cal W}_b\equiv\frac{|s_b|}{\gamma_{b2}}
&=3\sqrt2\,\eta\,
\frac{\chi^{3/2}|r-1|}{(r+\chi)^{5/2}},
\end{aligned}
\label{eq:shapeCompetition}
\end{equation}
where $\eta\equiv n_aZ_a^2/(n_bZ_b^2)$.
At fixed $r\ne1$, both scales vanish at extreme mass
ratios and peak at $\chi=2r/3$ and $3r/2$, respectively.
Thus this lowest-order deformation competes most effectively
with self-thermalization when the thermal speeds are comparable,
$T_a/m_a\sim T_b/m_b$.
Increasing $\eta$ weakens this competition for species
$a$ and strengthens it for $b$, which means the relatively dilute or lower-charge species is more susceptible to deformation.
Density and charge ratios therefore redistribute the shape response between the species. We note that these scales characterize the lowest-order shape response; higher-order distortions can contribute significantly at extreme species parameters.

The accumulated delay also depends on the time spent at different temperature ratios.
As the temperatures become comparable, energy exchange slows with the remaining temperature difference [Eq.~(\ref{eq:spitzer})], giving this stage substantial time weight even for widely separated initial temperatures.
Representative FP histories support this picture: most of
the delay accumulates during this stage (see End Matter).
The thermal-speed matching identified above then favors
comparable masses, suggesting enhanced accumulated feedback at intermediate mass ratios.

\begin{figure}[h]
\includegraphics[width=0.98\columnwidth]{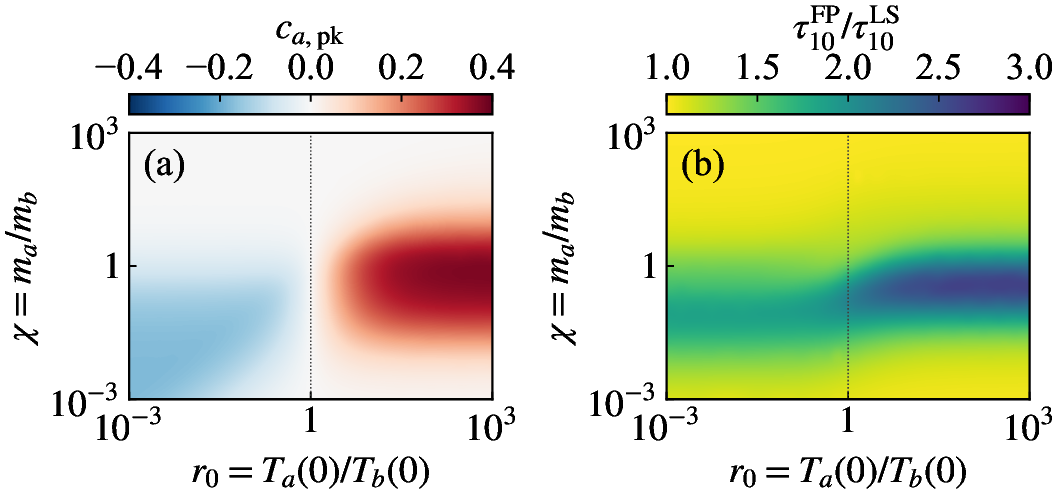}
\caption{Species dependence of distribution deformation and relaxation delay.
(a) Signed peak fourth-moment departure $c_{a,\mathrm{pk}}$.
(b) Relaxation-time ratio $\tau_{10}^{\mathrm{FP}}/\tau_{10}^{\mathrm{LS}}$.
Both are plotted against the initial temperature ratio
$r_0=T_a(0)/T_b(0)$ and mass ratio $\chi=m_a/m_b$.
FP results are shown for $n_a=n_b$, $Z_a/Z_b=1/2$,
and equal, fixed Coulomb logarithms.}
\label{fig:cnF}
\end{figure}

Figure~\ref{fig:cnF} maps the parameter dependence of
this moment and the relaxation delay, both obtained
from full FP simulations, at fixed density and charge ratios.
The signed peak
$c_{a,\mathrm{pk}}\equiv c_a(t_{\max|c_a|})$
is negative for the heated species and positive for the
cooled species [Fig.~\ref{fig:cnF}(a)].
Large distortions and delays occur at intermediate mass
ratios, while their maxima do not coincide because the
delay integrates the feedback from all modes of both distributions over time.
Although the initial feedback can enhance energy exchange
at extreme temperature ratios, the relaxation-time ratio
exceeds unity throughout the scan, supporting the substantial
time weight of the subsequent comparable-temperature stage.
The maximum ratio in the scan
$\tau_{10}^{\mathrm{FP}}/\tau_{10}^{\mathrm{LS}}\simeq2.6$ occurs at $\chi\approx0.316$
[Fig.~\ref{fig:cnF}(b)].

\textit{Consequences in fusion}---In inertial confinement fusion (ICF), shocks can heat different ion species unequally \cite{Rinderknecht2015}, providing a setting to test our predictions. In the strong-shock limit, mass-proportional ion heating gives $T_{\mathrm T}/T_{\mathrm D}\simeq m_{\mathrm T}/m_{\mathrm D}=1.5$ for DT fuel \cite{Keenan2018,Kabadi2021}. For the ion-only model, full FP calculations give \(\tau_{10}^{\mathrm{FP}}/\tau_{10}^{\mathrm{LS}}\simeq1.51\). To include the effect of electrons, we evolve all three distributions
in equimolar DT at $\rho=100\,\mathrm{g\,cm^{-3}}$, with
$T_{\mathrm D}(0)=4$ keV, $T_{\mathrm T}(0)=6$ keV,
$T_e(0)=5$ keV, and
$n_e=n_{\mathrm D}+n_{\mathrm T}$.
Using pair-dependent Coulomb logarithms~\cite{Beresnyak2023} updated at the instantaneous temperatures, the D--T relaxation time remains approximately 47\% longer than the corresponding three-species LS prediction. For these parameters, the corresponding relaxation times are approximately 0.17 ps for LS and 0.25 ps for FP.

\begin{figure}[h]
\centering
\includegraphics[width=0.98\columnwidth]{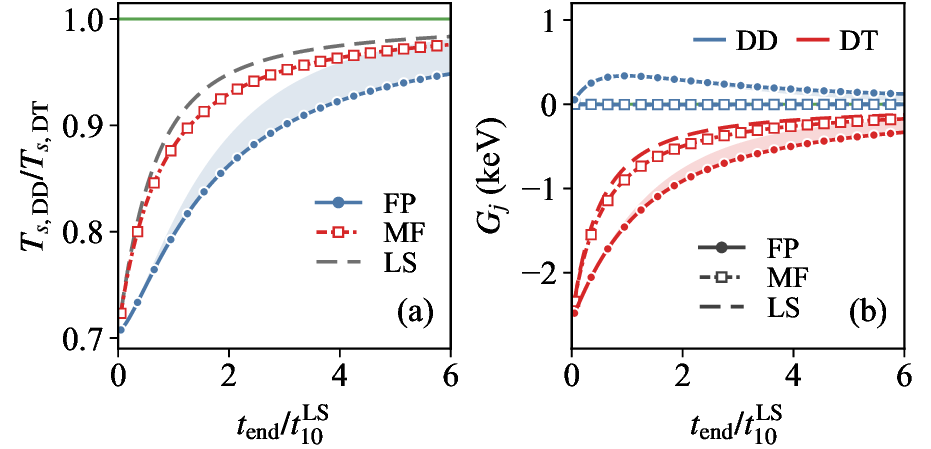}
\caption{Neutron spectral signatures of non-Maxwellian relaxation.
(a) Burn-averaged apparent temperature ratio $T_{s,\mathrm{DD}}/T_{s,\mathrm{DT}}$.
(b) Gamow-shift residuals $G_j$ for DD (blue) and DT (red).
Curves show constant-density FP, MF, and LS results;
MF retains the FP temperatures with Maxwellian distributions.
The ion-only plasma is equimolar with $T_{\mathrm D}(0)=4$~keV
and $T_{\mathrm T}(0)=6$~keV.
Shading spans FP results under adiabatic compression,
$C(x)=1+(C_6-1)x/6$, where $C=n/n_0$,
$x=t/t_{10}^{\mathrm{LS}}$, $1\leq C_6\leq2$,
and $\bar T\equiv(T_{\mathrm D}+T_{\mathrm T})/2=5C^{2/3}$~keV.
Spectra are integrated from $0$ to $t_{\mathrm{end}}$
over a fixed-mass fluid element in its local rest frame.}
\label{fig:man}
\end{figure}

Besides the temperature discrepancy, the non-Maxwellian distortions also affect fusion observables. Relative to Maxwellians at the same instantaneous temperatures, the cooled T species develops an enhanced high-energy tail, whereas the heated D species develops a depleted tail, as in the D--$\alpha$ example in Figs.~\ref{fig:num}(a) and \ref{fig:num}(b). Since the tail population contributes most to fusion reactivity, these distortions modify both reaction rates and neutron spectra~\cite{Appelbe2023,Appelbe2024}. To isolate the effect of these ion-distribution distortions, we return to the ion-only equimolar DT case with $T_{\mathrm D}(0)=4$~keV and $T_{\mathrm T}(0)=6$~keV, and accumulate the DD and DT neutron spectra~\cite{Munro2016} from $t=0$ to $t_{\mathrm{end}}$ using the Bosch--Hale cross sections \cite{Bosch_1992}. At $t_{\mathrm{end}}=t_{10}^{\mathrm{LS}}$, the cumulative DD and DT
yields are respectively 8.6\% and 2.2\% below the LS predictions.
Figure~\ref{fig:man} further compares the apparent temperature ratio $T_{s,\mathrm{DD}}/T_{s,\mathrm{DT}}$ and Gamow-shift residuals $G_j=\bar E_j-E_{M,j}(T_{s,j})$ for LS, FP, and Maxwellians following the FP temperature history (MF). The spectral temperature is defined by $V_{M,j}(T_{s,j})=V_j$, where $j$ labels the DD or DT reaction, $\bar E_j$ and $V_j$ are the mean and variance of the cumulative neutron spectrum, and $E_{M,j}(T)$ and $V_{M,j}(T)$ are their counterparts for Maxwellian
reactants at a common temperature $T$ \cite{Crilly2022}.  While temperature separation already lowers $T_{s,\mathrm{DD}}/T_{s,\mathrm{DT}}$~\cite{Inglebert2014},
non-Maxwellian distortions lower it further relative to MF [Fig.~\ref{fig:man}(a)]. The FP spectra also exhibit positive DD and negative DT Gamow-shift residuals [Fig.~\ref{fig:man}(b)]. Physically, these opposite residuals are consistent with the departures produced by enhanced and truncated high-energy tails in Fig.~5 of Ref.~\cite{hartouni2023evidence}.

We also briefly consider the effect of compression. For homogeneous isotropic adiabatic compression, the scalings $n_s\to Cn_s$, $T_s\to C^{2/3}T_s$, and $v\to C^{1/3}v$ preserve the distribution shapes $g_s$ and the collision frequencies $\nu_{ab}\propto n_b/T_a^{3/2}$ for fixed Coulomb logarithms.
Thus, the collisional rate correction $F$ and the relaxation-time ratio remain unchanged when the temperature difference is scaled by the evolving mean temperature.
Compression modifies the neutron observables through the higher absolute temperatures and burn weighting. As shown by the shaded regions in Fig.~\ref{fig:man}(b), the increased reaction weight of the hotter, later plasma reduces the residual magnitudes at later stopping times while preserving their signs over the histories examined. More generally, the visibility of these signatures depends on how strongly neutron production weights the relaxation transient.

These opposite DD and DT residuals may be relevant to the high-temperature shock-driven implosions in Ref.~\cite{Mannion2023}, where DD measurements lie consistently above the Maxwellian curve ($G_{\mathrm{DD}}>0$) while DT measurements lie both above and below it. Alongside counterstreaming shock-precursor ions~\cite{Mannion2023}, hydrodynamic profiles, residual flows~\cite{Crilly2022}, and kinetic tail depletion~\cite{Molvig2012,Kagan2015}, intrinsic multi-ion collisional relaxation may also contribute to these diagnostic signatures.

A related process may become increasingly important as ICF advances toward higher fuel burn fractions, where fusion-produced $\alpha$ particles sustain fuel heating and further fusion~\cite{Zylstra2022}. 
Born highly energetic and nonthermal, these particles lose substantial energy to electrons before predominantly heating the fuel ions and relaxing toward Maxwellian distributions~\cite{Peigney2014}.
The late ion-heating stage may therefore involve collisional shape feedback similar to that identified here, modifying both energy deposition into the fuel and its subsequent fusion reactivity.

Indeed, simulations of nonequilibrium burn have shown fuel distortions qualitatively resembling the ``$-+-$'' pattern (see Fig.~4 of Ref.~\cite{Hu2025}) and slower heating than a model that constrains D and T to remain Maxwellian, suggesting a possible shared physical origin.
If $\alpha$ heating produces similar tail depletion in both
fuel species, positive Gamow-shift residuals could arise in
both DD and DT spectra. This possibility suggests a connection with the DT spectral upshifts observed in burning plasmas~\cite{hartouni2023evidence}.

\textit{Conclusion and discussion}---Temperature relaxation is itself a source of non-Maxwellianity: the same interspecies collisions that exchange energy also deform the distributions and modify the exchange rate. Full FP calculations show that these distortions can accumulate into substantial relaxation delays, with their parameter dependence reflecting the competition between interspecies deformation and self-thermalization. A temperature-only description can therefore miss essential kinetic physics even in a homogeneous isotropic plasma.

In ICF, delayed equilibration prolongs ion-temperature separation, while the accompanying tail distortions modify fusion reactivity and neutron spectra. Extending this framework to couple $\alpha$-particle production, slowing, and thermalization with the evolving D, T, and electron distributions will clarify how collisional shape feedback affects fuel heating and burn.

Similar effects may also arise in unequal-temperature astrophysical plasmas, including solar-wind proton--$\alpha$ mixtures \cite{Maruca2013,Johnson2023} and the downstream plasma of supernova-remnant shocks \cite{Miceli2019}, potentially affecting ion-temperature evolution and the interpretation of spectroscopic diagnostics.

\textit{Acknowledgments}---Yao Guo thanks Dr.\ Jiong-Hang Liang for useful discussions. This work was supported by Science Challenge Project (No.\ TZ2025012), the Strategic Priority Research Program of the Chinese Academy of Sciences (Grant Nos.\ XDA25010100 and XDA25050500), National Natural Science Foundation of China (Grant No.\ 12075204), and Shanghai Municipal Science and Technology Key Project (Grant No.\ 22JC1401500). Dong Wu thanks the sponsorship from Yangyang Development Fund.


\bibliography{apssamp}

\clearpage

\appendix*

\section{End Matter}
\setcounter{equation}{0}
\renewcommand{\theequation}{A\arabic{equation}}

\subsection{Three-lobe structure of $\partial_t g|_M$}
With $x=\alpha u_a>0$, Eqs.~(\ref{eq:spitzer}) and (\ref{eq:MaxwellianDeparture}) give
\begin{align}
\frac{\left.\partial_t g_a\right|_M}
     {\nu_{ab}g_M}
&=2\alpha\chi\left(1-\frac{1}{r}\right)
G_\alpha(x),\label{eq:pgm}\\
G_\alpha(x)&=
\frac{2(1+\alpha^2)}{\sqrt{\pi}}e^{-x^2}
-\frac{\operatorname{erf}(x)}{x}
+\frac{4x^2-6\alpha^2}
 {3\sqrt{\pi}(1+\alpha^2)^{3/2}}.
\nonumber
\end{align}
$G_\alpha(0)=(2\alpha^2/\sqrt{\pi})
\left[1-(1+\alpha^2)^{-3/2}\right]>0$, and
$G_\alpha(x\rightarrow\infty)\sim 4x^2/[3\sqrt{\pi}(1+\alpha^2)^{3/2}]>0$.
Particle conservation requires $G_\alpha<0$ somewhere, so it has at least two positive zeros.

$xG_\alpha(x)$ shares the sign with $G_\alpha(x)$. Its extrema satisfy
\begin{equation}
[2(1+\alpha^2)y-\alpha^2]e^{-y}
=
\frac{2y-\alpha^2}{(1+\alpha^2)^{3/2}},
\qquad y=x^2.
\label{eq:Gstationary}
\end{equation}
Solutions are possible only for
$0<y<\alpha^2/[2(1+\alpha^2)]$ or $y>\alpha^2/2$.
In either interval, the logarithm of the ratio of the two sides
of Eq.~(\ref{eq:Gstationary}) has derivative
\begin{equation}
-1-\frac{2\alpha^4}
{[2(1+\alpha^2)y-\alpha^2](2y-\alpha^2)}<0.
\end{equation}
The logarithmic ratio runs from $(3/2)\ln(1+\alpha^2)$
to $-\infty$ in the first interval and from $+\infty$ to
$-\infty$ in the second. Thus $xG_\alpha$ has one maximum
and one minimum. Since it starts at zero with positive slope,
$xG_\alpha$ has exactly two positive zeros and so does $G_\alpha$, giving $+\,-\,+$
for $T_a>T_b$ and $-\,+\,-$ for $T_a<T_b$.

\subsection{Exact initial feedback on the relaxation rate}
At a double-Maxwellian state, $F_a=0$, and the shape variation of the energy moment is
\begin{align}
\delta Q_a
={}&\int_0^\infty4\pi u^2\delta g_a(u)
\epsilon_a(u,g_M)\,\d u
\nonumber\\
&+\int_0^\infty4\pi u^2g_M(u)
\delta\epsilon_a[\delta g_b](u)\,\d u.
\label{eq:deltaQepsilonEnd}
\end{align}
Interchanging the integration order in the second term gives
\begin{equation}
\int_0^\infty u^2g_M\,
\delta\epsilon_a[\delta g_b]\,\d u
=-\alpha\chi\frac{\nu_{ab}}{\nu_{ba}}
\int_0^\infty u^2\delta g_b
\epsilon_b(u,g_M)\,\d u.
\label{eq:fieldVariationEpsilonEnd}
\end{equation}
Setting $\delta g_s=\partial_tg_s|_M\,\delta t$
then yields Eq.~(\ref{eq:exactFVariationEnd}), with
\begin{align}
{\cal P}(\alpha,\chi)=\frac{4}{\sqrt{\pi}}\bigg[
&\frac{\sqrt{1+\alpha^2}}{\alpha}
\frac{\chi(\alpha^4+3\alpha^2+1)-\alpha^4}
{(1+2\alpha^2)^{3/2}}
\nonumber\\
&-\frac{\chi(1+\alpha^2)^{3/2}}{\alpha^3}
\tan^{-1}\!\left(
\frac{\alpha^2}{\sqrt{1+2\alpha^2}}\right)
\nonumber\\
&-\frac{\alpha^3(\chi\alpha^2-2\chi-3)}
{3(1+\alpha^2)^{5/2}}
\bigg].
\label{eq:exactPEnd}
\end{align}
Consequently,
\begin{equation}
\frac{\dot F_a|_M}{\nu_{ab}}
={\cal P}(\alpha,\chi)
+\frac{n_a}{n_b}\chi^2\alpha^3
{\cal P}(\alpha^{-1},\chi^{-1}).
\label{eq:exactSignCriterionEnd}
\end{equation}
The charge states therefore do not affect the sign.

Label the heavy species $a$ and the light species $b$, so $\chi\geq1$ and $\alpha=\alpha_{hl}$. Write
${\cal P}(\alpha,\chi)
=\chi{\cal P}_1(\alpha)+{\cal P}_2(\alpha)$.
Equation~(\ref{eq:exactPEnd}) gives
\begin{equation}
{\cal P}_2(\alpha)=\frac{4\alpha^3}{\sqrt{\pi}}
\left[
\frac{1}{(1+\alpha^2)^{5/2}}
-\frac{\sqrt{1+\alpha^2}}{(1+2\alpha^2)^{3/2}}
\right]<0,
\end{equation}
since $(1+\alpha^2)^2>1+2\alpha^2$.
Set $x=\alpha^2$ and
$H(x)\equiv\sqrt{\pi}x^{3/2}{\cal P}(\sqrt{x},1)/
[4(1+x)^{3/2}]$.
Then $H(0)=0$ and
\begin{equation}
\begin{aligned}
H'(x)&=-\frac{x^2}{(1+x)^5}
\left[
\frac{(5+7x)(1+x)^3}{(1+2x)^{5/2}}+3x-5
\right]\\
&<-\frac{6x^4}{(1+x)^5(1+2x)}<0.
\end{aligned}
\end{equation}
Thus ${\cal P}(\alpha,1)<0$. The light-species term is
${\cal P}(\alpha^{-1},\chi^{-1})
=\chi^{-1}{\cal P}(\alpha^{-1},1)
+(1-\chi^{-1}){\cal P}_2(\alpha^{-1})<0$.

Since ${\cal P}_1\sim-24\alpha^7/(5\sqrt{\pi})$ near zero
and its first positive zero is $\alpha_c\simeq3.699$,
the heavy-species term is also negative for $\alpha<\alpha_c$.
Positive initial feedback therefore requires,  at a minimum,
$\alpha_{hl}>\alpha_c$.
For equal densities, its sign boundary has minimum
$T_h/T_l\simeq126.6$.

\begin{figure}[htbp]
\centering
\includegraphics[width=0.98\columnwidth]{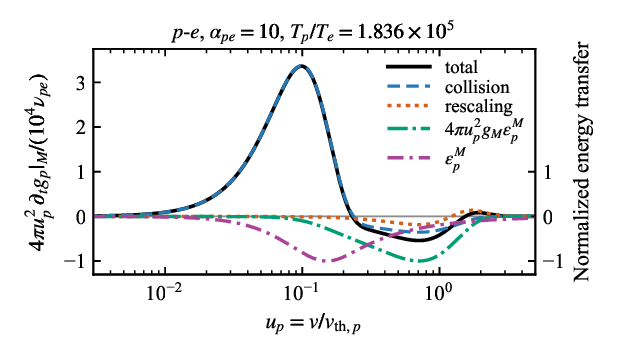}
\caption{Enhanced initial energy exchange in the hot-heavy regime.
Left axis: proton deformation source (black), with contributions
from interspecies collisions (blue dashed) and thermal rescaling (orange dotted).
Right axis: energy transfer rate per unit $u_p$, $4\pi u_p^2g_M\epsilon_p^M$ (green)
and per-particle energy transfer $\epsilon_p^M$ (purple),
each normalized by its maximum magnitude.
Parameters are $n_p=n_e$, $m_p/m_e=1836$,
and $T_p/T_e=1.836\times10^5$.}
\label{fig:hotProtonEnd}
\end{figure}

In the extreme hot-heavy regime, per-particle cooling efficiency $|\epsilon_h^M|$ peaks
at $u_h=O(\alpha_{hl}^{-1})$, far below the peak of cooling rate per unit $u_h$, $4\pi u_h^2g_M|\epsilon_h^M|$. Redistribution toward low velocities can then enhance
cooling. Figure~\ref{fig:hotProtonEnd} illustrates this for
$\alpha_{pe}=10$: the two peaks lie at $u_p\simeq0.151$
and $0.707$, and both species together give
$\dot F|_M/\nu_{pe}\simeq6.62\times10^3>0$.

\subsection{Lowest-order Sonine closure}
\begin{figure}[htbp]
\includegraphics[width=0.98\columnwidth]{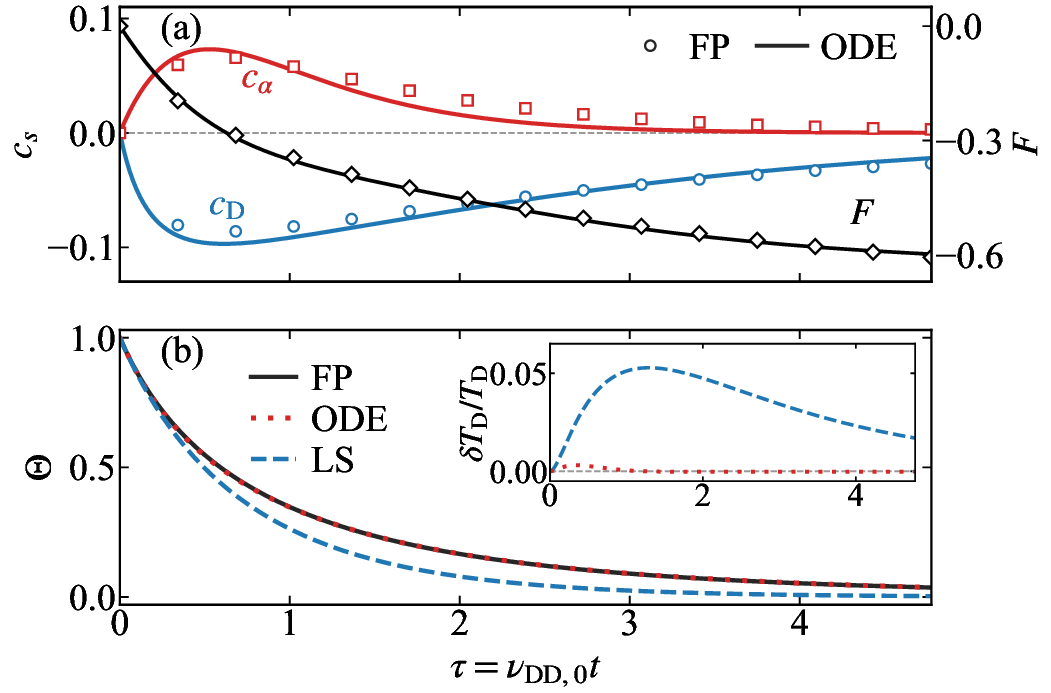}
\caption{Lowest-order Sonine closure of temperature relaxation.
(a) Fourth-moment departures $c_s$ and rate correction $F$
from FP (markers) and the closure ODEs (solid).
(b) Normalized temperature difference $\Theta=[T_\alpha(t)-T_{\mathrm D}(t)]/
[T_\alpha(0)-T_{\mathrm D}(0)]$: FP, ODE, and LS.
Inset: $(T_{\mathrm D}^{\mathrm{model}}-T_{\mathrm D}^{\mathrm{FP}})/
T_{\mathrm D}^{\mathrm{FP}}$.
Parameters are those of the D--$\alpha$ plasma in Fig.~\ref{fig:num};
time is normalized as $\tau=\nu_{\mathrm{DD},0}t$.}
\label{fig:sonineEnd}
\end{figure}
The lowest-order nontrivial Sonine basis function,
$L_2^{(1/2)}(u^2)g_M$, is also the only one with a three-lobe
structure. We therefore examine to what extent this mode alone captures the temperature evolution. We retain $g_s\simeq g_M+c_s\phi_2$,
where
$\phi_2\equiv L_2g_M$ and
$L_2(u)\equiv L_2^{(1/2)}(u^2)
=u^4/2-5u^2/2+15/8$.
This perturbation preserves density and temperature.
Keeping terms linear in $c_s$ gives the interspecies term
\begin{equation}
\begin{aligned}
{\cal K}_{ab}\equiv{}&\hat C_{ab}[g_M,g_M]
+c_a\hat C_{ab}[\phi_2,g_M]\\
&+c_b\hat C_{ab}[g_M,\phi_2].
\end{aligned}
\end{equation}
Projection onto $L_2$ and the energy moment gives, for $s=a,b$,
\begin{equation}
\begin{aligned}
N_2\left(\dot c_s+2\frac{\dot T_s}{T_s}c_s\right)
&=S_s+\sum_{t=a,b}M_{st}c_t,\\
\frac{3\dot T_s}{2T_s}
&=Q_s^M+\sum_{t=a,b}R_{st}c_t,
\end{aligned}
\label{eq:sonineClosureEnd}
\end{equation}
where $N_2=15/8$. Here $S_s$, $M_{st}$, and $R_{st}$
are the shape source, shape-response coefficients, and
energy-response coefficients, respectively.
Writing $(A,B)\equiv\int_0^\infty4\pi u^2AB\,\d u$ and
$\hat H_{s2}\equiv\hat C_{ss}[\phi_2,g_M]
+\hat C_{ss}[g_M,\phi_2]$, their definitions are
\begin{equation}
\begin{aligned}
S_a&=(L_2,\hat C_{ab}[g_M,g_M]),\\
M_{at}&=(L_2,\partial_{c_t}{\cal K}_{ab}
+\delta_{at}\hat H_{a2}),\\
R_{at}&=(u^2,\partial_{c_t}{\cal K}_{ab}),
\end{aligned}
\end{equation}
with derivatives at fixed temperatures and $\delta_{at}$
the Kronecker delta. Exchange of species gives the remaining
coefficients, closing four ODEs for $T_a,T_b,c_a,c_b$.

The source $s_a=S_a/N_2$ and projected self-collision rate
entering Eq.~(\ref{eq:shapeCompetition}) are
\begin{equation}
\begin{aligned}
s_a&=\frac{16\nu_{ab}}{5\sqrt{\pi}}(r-1)
\frac{\alpha^3}{(1+\alpha^2)^{5/2}},\\
\gamma_{s2}&=-\frac{(L_2,\hat H_{s2})}{N_2}
=\frac8{15}\sqrt{\frac2\pi}\,\nu_{ss}.
\end{aligned}
\end{equation}
Using $\alpha^2=r/\chi$ and $\nu_{ab}/\nu_{aa}=\eta^{-1}$
gives ${\cal W}_a$; exchange of species gives ${\cal W}_b$.

For the initially Maxwellian D--$\alpha$ case in
Fig.~\ref{fig:num}, we evolve these equations from the same
initial temperatures with $c_s=0$.
The closure reproduces the FP temperature history, giving
$\tau_{10}^{\mathrm{ODE}}=2.825$, approximately 0.4\% above the FP value,
despite larger discrepancies in the shape amplitudes
[Fig.~\ref{fig:sonineEnd}(a)]. Our scan suggests that the lowest-order closure is useful for moderate species parameters but can break down at extreme mass, temperature, or density ratios.
Its accuracy for comparable thermal speeds ($\alpha\sim1$) is consistent with a simple estimate from the Sonine projections, which indicates that higher-order contributions to energy exchange are suppressed.

\subsection{Fokker--Planck calculations}
We evolve the coupled isotropic FP equations for all species
in physical velocity coordinates, equivalent to
Eq.~(\ref{eq:FP2}), using a semi-implicit finite-volume scheme
with Chang--Cooper fluxes~\cite{Chang1970} and zero-flux
velocity boundaries.
Temperatures evolve through the energy moments
[Eq.~(\ref{eq:Srate})], with pairwise energy transfer
antisymmetrized to conserve total energy.
For the three-species case, the collision terms and energy
moments are summed over collision partners.
All distributions are initially Maxwellian.
The LS reference histories follow Eq.~(\ref{eq:spitzer})
with the same initial conditions and Coulomb logarithm
prescription.
Representative relaxation times are checked under
velocity-grid and time-step refinement; for the
three-species DT case, these refinements change the
relaxation-time ratio by less than 0.02\%.

Integrating $-F$ along the two-species FP histories
quantifies the contribution of each temperature interval
to the accumulated delay.
At $(r_0,\chi)=(10^{-3},0.1)$ and $(10^3,10^{-1/2})$,
respectively, 81\% and 74\% of this delay accumulate within
$1/2<T_a/T_b<2$, supporting the substantial contribution
of the comparable-temperature stage.

\end{document}